\documentclass[twocolumn,superscriptaddress,longbibliography,aps,pra,preprintnumbers]{revtex4-2}

\usepackage{graphicx}
\usepackage{bm}
\usepackage{color}
\usepackage{epstopdf}
\usepackage{amsmath}
\usepackage{amssymb}
\usepackage{epstopdf}
\usepackage{gensymb}
\usepackage{float}

\usepackage[urlcolor=blue,colorlinks=true,citecolor=blue,linkcolor=blue,pdfstartview={FitH},bookmarks=false]{hyperref}

\usepackage{ulem}
\definecolor{sdgreen}{rgb}{0.0, 0.5, 0.0}
\definecolor{myblue}{rgb}{0.0, 0.5, 0.3}

\definecolor{pjorange}{rgb}{0.8, 0.3, 0.0}

\begin{document}

\title{Anharmonicity and structural phase transition in the Mott insulator Cu$_2$P$_2$O$_7$}

\author{Svitlana Pastukh}
\email[corresponding author; e-mail: ]{svitlana.pastukh@ifj.edu.pl}
\affiliation{Institute of Nuclear Physics, Polish Academy of Sciences, PL-31342 Krak\'{o}w, Poland}

\author{Pawe\l{} T. Jochym}
\affiliation{Institute of Nuclear Physics, Polish Academy of Sciences, PL-31342 Krak\'{o}w, Poland}

\author{Oleksandr Pastukh}
\affiliation{Institute of Nuclear Physics, Polish Academy of Sciences, PL-31342 Krak\'{o}w, Poland}

\author{Jan \L{}a\.{z}ewski}
\affiliation{Institute of Nuclear Physics, Polish Academy of Sciences, PL-31342 Krak\'{o}w, Poland}

\author{Dominik Legut}
\affiliation{IT4Innovations, VSB - Technical University of Ostrava, 70800 Ostrava, Czech Republic}
\affiliation{Department of Condensed Matter Physics, Faculty of Mathematics and Physics, Charles University, 121 16 Prague, Czech Republic}

%\affiliation{Nanotechnology Centre, CEET VSB - Technical University of Ostrava, 70800 Ostrava, Czech Republic}

\author{Przemys\l{}aw Piekarz}
%\email[e-mail: ]{przemyslaw.piekarz@ifj.edu.pl}
\affiliation{Institute of Nuclear Physics, Polish Academy of Sciences, PL-31342 Krak\'{o}w, Poland}

\date{\today}

\begin{abstract}
{\it Ab initio} investigations of structural, electronic, and dynamical properties of the high-temperature $\beta$ phase of copper pyrophosphate were performed using density functional theory. 
The electronic band structure shows the Mott insulating state due to electron correlations in copper ions.  By calculating phonon dispersion relations, the soft mode at the A point of the Brillouin zone was revealed, showing the dynamical instability of the $\beta$ phase at low temperatures. The double-well potential connected with the soft mode is derived and the mechanism of the structural phase transition to the $\alpha$ phase is discussed. The self-consistent phonon calculations based on the temperature-dependent effective potential show the stabilization of the $\beta$ phase at high temperatures, due to the anharmonic effects. 
The pronounced temperature dependence and the large line width of the soft mode indicate an essential role of anharmonicity in the structural phase transition.
\end{abstract}

\maketitle

\section{Introduction}

Transition metal pyrophosphates gained significant interest as promising materials for energy storage devices~\cite{Huang2015},
solid electrolytes~\cite{Venckute2015}, supercapacitors~\cite{Patil2023}, and photonics~\cite{Laskowska2020}.   
Copper pyrophosphate (Cu$_2$P$_2$O$_7$) has been extensively studied due to its magnetic, photoluminescence, dielectric, electrochemical, and catalytic properties~\cite{Stiles1972,Stiles1973,Janson2011,Agarwal2021,In-noi2021,Sang2022,Akiyama2022}.
The X-ray diffraction measurements on Cu$_2$P$_2$O$_7$ revealed the structural phase transition~\cite{Robertson1967} 
and the strong negative thermal expansion (NTE) in a broad temperature range 0--375~K~\cite{Shi2020}. 
Both these phenomena are related to the same vibrations of oxygen atoms, but the details of their mechanisms are not completely understood.

\begin{figure}[t!]
  \centering
    	\includegraphics[width=7.5cm]{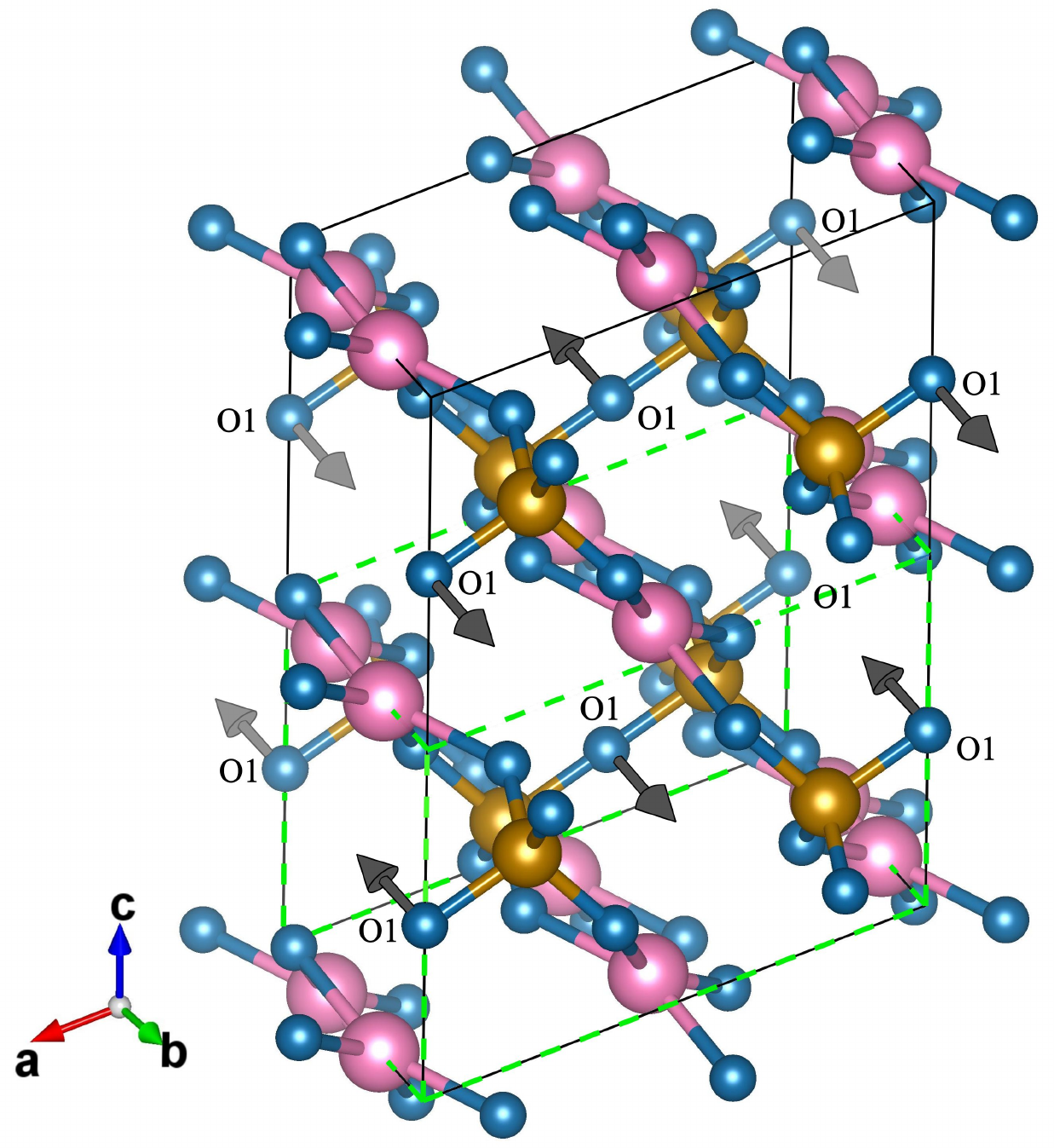}
    \caption{Crystal structure of Cu$_2$P$_2$O$_7$. The unit cells of the $\alpha$ and $\beta$ phases are plotted by the solid black and dashed green lines, respectively. 
    The following color scheme is used: Cu -- pink, P -- yellow, and O -- blue. The black arrows indicate oxygen displacements in the phonon soft mode connected with the $\alpha - \beta$ phase transition.
    The image was rendered using VESTA software~\cite{momma2011vesta}.}
  \label{fig:structure}
\end{figure}

Copper pyrophosphate crystallizes in two different polymorphic forms: low-temperature $\alpha$ and high-temperature $\beta$ phase,
and the reversible phase transition occurs around $T=363$~K~\cite{Pogorzelec2006}. As it was found by diffraction studies, \mbox{$\alpha$-Cu$_2$P$_2$O$_7$} has the monoclinic $C2/c$ space group with Cu$^{2+}$ cations in irregular octahedral coordination and the [PO$_4$]$^{3-}$ tetrahedra sharing one corner with the bridging oxygen atom O1 (see Fig.~\ref{fig:structure})~\cite{Robertson1967}. At higher temperatures, the enhanced thermal motion of O1 perpendicular to the P--O--P bonding makes it positionally disordered and induces transformation to the $\beta$ phase, which exhibits the monoclinic $C2/m$ structure~\cite{Robertson1967,Robertson1968}.
Diffuse scattering measurements detected an intermediate state in the temperature range of 347--363~K~\cite{Pogorzelec2006}. Furthermore, the $\alpha$--$\beta$ structural transformation was classified as a second-order phase transition of an inhomogeneous type by Robertson and Calvo~\cite{Robertson1967}.

Cu$_2$P$_2$O$_7$ exhibits the long-range antiferromagnetic (AFM) order below the N\'{e}el temperature $T_\text{N}=27$~K~\cite{Stiles1972,Stiles1973}. A two-dimensional model with strongly coupled magnetic dimers was investigated on the basis of the density functional theory (DFT) calculations~\cite{Janson2011}. 
In our previous studies on $\alpha$-Cu$_2$P$_2$O$_7$, we demonstrated that local Coulomb interactions in the Cu($3d$) states are responsible for the Mott insulating state with the AFM order~\cite{Pastukh2021}. We analyzed the lattice's dynamic properties by calculating the phonon spectrum for the optimized structure.
Recently, our findings were confirmed by similar results obtained using the hybrid-DFT calculations~\cite{Yang2023}. 

The aim of the present first-principles study is to obtain detailed information on the structural, electronic, and dynamical properties of $\beta$-Cu$_2$P$_2$O$_7$, and to explain the mechanism of the $\alpha$--$\beta$ phase transition. 
The calculations of the phonon dispersion relations revealed the phonon soft mode, which is responsible for the structural transformation.
We also demonstrate a crucial role of anharmonicity in the stabilization of the $\beta$ phase at higher temperatures.

\section{Calculation Details}

The calculations were performed using the projector augmented-wave method~\cite{Blochl1994} and the generalized gradient approximation (GGA)~\cite{PBE} implemented in the Vienna Ab initio Simulation Package~\cite{Kresse1996}. The integration in the $k$-point space was performed over the $4 \times 4 \times 8$ Monkhorst-Pack mesh \cite{MP} with the cut-off energy equal to 500 eV. The ground state energy for the AFM order was found to be consistent with the experimental studies~\cite{Stiles1972,Stiles1973} and previous calculations for the $\alpha$ phase~\cite{Janson2011,Pastukh2021,Yang2023}. During our investigation of the $\alpha$ phase, we utilized the GGA method with $U=J=0$, to reveal a weak metallic state due to persistent electronic states at the Fermi energy. To improve this, we applied the GGA+U method ~\cite{GGAU}, using specific parameters: $U = 4$ eV, $J = 1$ eV for Cu(3d) states and $U = 2$ eV, $J = 0.5$ eV for O(2p) states.  This led us to identify an insulating state with the electronic band gap of $\Delta E = 1.66$ eV. However, it's noteworthy that this gap is too small in comparison to values reported for copper pyrophosphate materials~\cite{karaphun2018influence,Majtyka2022,Yang2023}. To refine our analysis, we applied larger values of the Coulomb parameter: $U=9$ eV~for copper and $U=4$~eV for oxygen. Such values are commonly used for superconducting cuprates. Additionally, these $U$ and $J$ values yielded good agreement with experimental lattice parameters, as observed in the $\alpha$ phase analysis~\cite{Pastukh2021}. We also included the van der Waals corrections during optimization of the crystal structure~\cite{D2,D2-VASP}.	

The interatomic force constants and phonon energies were obtained with the {\sc Alamode} software~\cite{Tadano2014}.
First, we used the direct method~\cite{Parlinski} to calculate the phonon dispersion relations and phonon density of states (PDOS) within the harmonic approximation. The Hellmann-Feynman forces were obtained by making small displacements of single atoms in the $2\times1\times2$ supercell with 88 atoms using the $2\times2\times2$ $k$-point grid for DFT. From this information, the force constants and dynamical matrices were constructed. The phonon frequencies and polarization vectors were calculated by the direct diagonalization of the dynamical matrix. The LO-TO splitting at the $\Gamma$ point is determined by calculating the static dielectric tensor and Born effective charges within the density functional perturbation theory~\cite{Gajdos2006}. This method was used previously to study lattice dynamical properties of  $\alpha$-Cu$_2$P$_2$O$_7$~\cite{Pastukh2021}.

In the present study, we utilize the temperature-dependent effective potential (TDEP) approach~\cite{Hellman2011,Hellman2013} to incorporate anharmonic effects. This method determines the atomic potential from forces induced by thermal atom displacements at finite temperatures.

\begin{figure}[t!]
  \centering
    	\includegraphics[width=9cm]{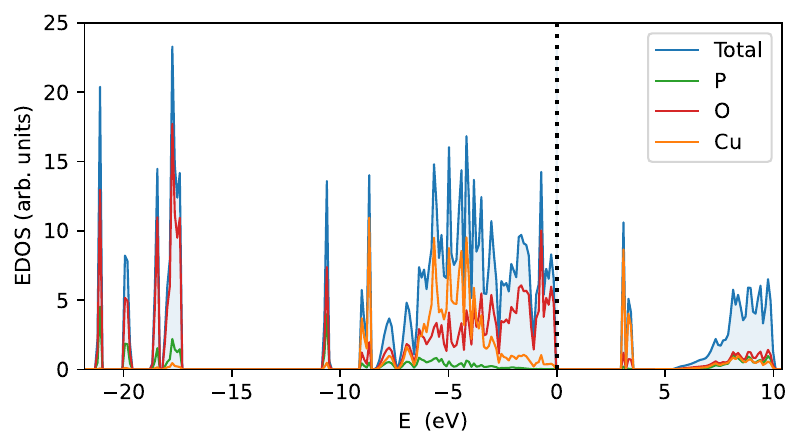}
    \caption{Total and partial electronic density of states for $\beta$ phase of Cu$_2$P$_2$O$_7$.}
  \label{fig:edos}
\end{figure}

The configurations of atomic displacements are generated by the high-efficiency configuration space sampling (HECSS) method~\cite{Jochym2021}, which was successfully applied to study the structural phase transition in the $A$V$_3$Sb$_5$ ($A$ = K, Rb, Cs) superconductors~\cite{Ptok2022} and the negative thermal expansion of $\alpha$-Sn~\cite{Jochym2022}.  
In order to investigate the effect of anharmonicity and the temperature dependence of phonon frequencies, we apply the self-consistent phonon (SCPH) method~\cite{Tadano2015}. Additionally, we calculate the frequency shift and line width of phonons using the real and imaginary parts of the phonon self-energy resulting from the quartic and cubic anharmonicity~\cite{Tadano2018}.

\section{CRYSTAL AND ELECTRONIC STRUCTURE}

The Cu$_2$P$_2$O$_7$ crystal structure consists of pyrophosphate P$_2$O$_7$ layers parallel to the (001) crystallographic plane and Cu cations coordinated to four oxygen atoms (Fig.~\ref{fig:structure}). The diphosphate anion [P$_2$O$_7$]$^{4-}$ is made up of two [PO$_4$]$^{3-}$ tetrahedra connected by a shared oxygen atom O1. This central oxygen atom in the P--O--P bridge exhibits the large mean-square displacement in the $\alpha$ phase~\cite{Shi2020} and positional disorder in $\beta$-Cu$_2$P$_2$O$_7$~\cite{Robertson1967,Robertson1968}.
After the optimization, we obtained the lattice parameters: $a=6.86$~\AA, $b=8.09$~\AA, $c=4.32$~\AA, and $\beta=109.6$\degree, in a good agreement with the experimental data~\cite{Pogorzelec2006}. 
The size of the unit cell along the $c$ direction in the $\beta$ phase is approximately two times smaller than in the $\alpha$ phase, and two other lattice constants $a$ and $b$ have similar values.

\begin{figure*}[t!]
  \centering
    	\includegraphics[width=18cm]{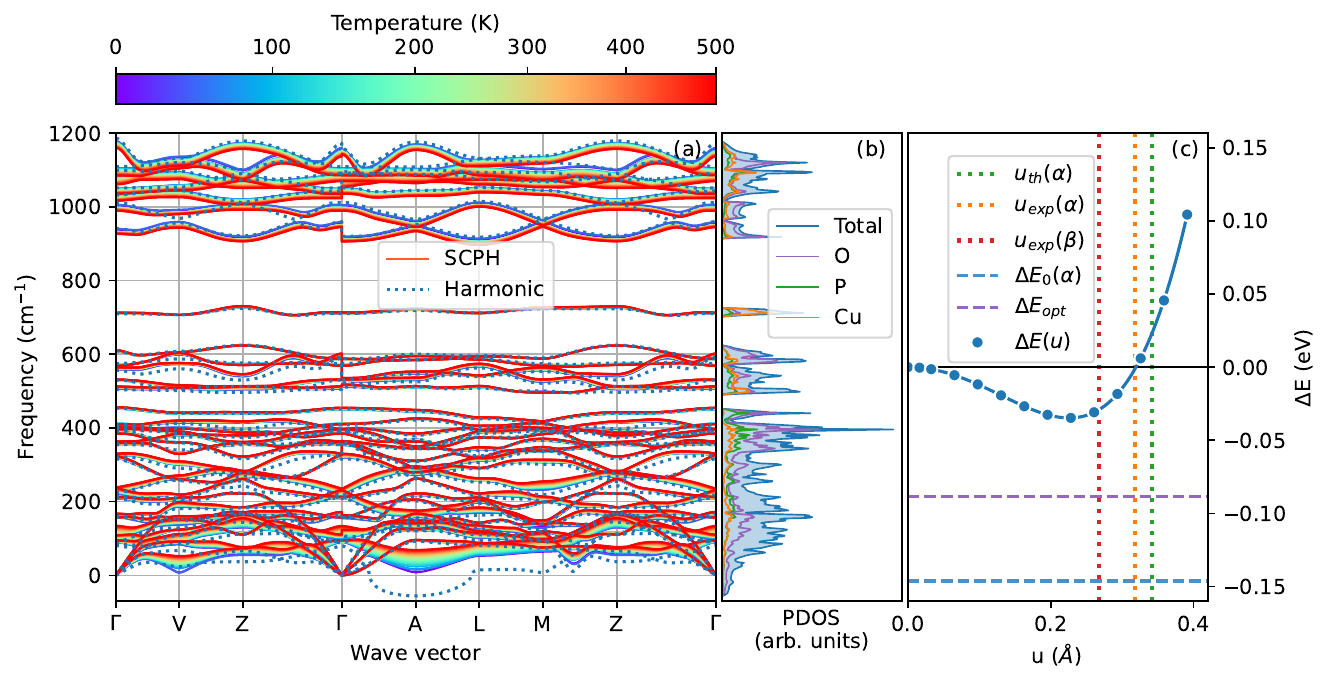}
    \caption{(a) Phonon dispersion relations and (b) phonon density of states obtained for $\beta$-Cu$_2$P$_2$O$_7$ within the harmonic approximation (dotted blue line) and using the TDEP+SCPH approach (rainbow color bar). (c) The energy potential of the soft mode at the A point plotted as the function of the amplitude of the O1 oxygen. 
    }
  \label{fig:phonons}
\end{figure*}

The total and partial element-specific electronic densities of states (EDOS) calculated for the AFM state in  $\beta$-Cu$_2$P$_2$O$_7$ is presented in Fig.~\ref{fig:edos}.
Close to the Fermi energy ($E_\text{F}$) the electronic states are dominated by the O($2p$) orbitals.
The lower Hubbard band corresponds to the peak around $-8$~eV, and it contains mainly the majority-spin Cu($3d$) states
with a small contribution from O($2p$) orbitals. The energy gap appears between the occupied states with both spin components 
and the upper Hubbard band with the empty minority-spin Cu($3d$) states.
The value of the energy gap $E_\text{g}=2.94$ eV is slightly reduced compared to the $\alpha$ phase~\cite{Pastukh2021}, and it corresponds well to the experimental value for the hydrated copper pyrophosphate (2.34~eV)~\cite{Majtyka2022}.

\section{LATTICE DYNAMICS AND STRUCTURAL PHASE TRANSITION}

The phonon dispersion relations of $\beta$-Cu$_2$P$_2$O$_7$ 
obtained within the direct and TDEP+SCPH methods are presented in Fig.~\ref{fig:phonons}(a).
In contrast to the $\alpha$ phase, which is stable at low temperatures~\cite{Pastukh2021},
the calculation within the harmonic approximation reveals the soft mode at the A point with the wave vector $\textbf{q}=(0,0,0.5)$.
In Fig.~\ref{fig:structure}, we present the displacements of the central oxygen atom O1, which show the largest amplitude in the soft mode. 
This oxygen atom vibrates along the $b$ direction, perpendicular to the P--O--P bridge, and the sign of its displacement changes 
in the following unit cells along the $c$ direction.
Therefore, when this soft mode condensates in the $\alpha$ phase, the unit cell is doubled along the $c$ direction.
As it was found previously \cite{Shi2020}, the O1 atoms show very large thermal displacements along the $b$ direction and their shifts from the central positions are responsible for the $\alpha$-$\beta$ transition~\cite{Robertson1967,Pogorzelec2006}. Here, we confirmed that vibrations of these atoms are connected with the soft mode, and therefore, when the temperature is decreased, these displacements along the $b$ direction determine the structure of the $\alpha$ phase.
Indeed, by the group theory analysis, we verified that the soft mode properly describes the change of symmetry, $C2/m (\beta) \rightarrow C2/c (\alpha)$.

In Fig.~\ref{fig:phonons}(b), we show the total PDOS and partial contributions of Cu, O, and P vibrations obtained within the direct method.
In the low-frequency range, the phonon states are dominated by Cu and O contributions. At intermediate frequencies ($\sim$300--600~cm$^{-1}$), the PDOS is dominated by vibrations of oxygen atoms, with a smaller contribution of P and Cu atoms. In the narrow band around 700~cm$^{-1}$ mainly the vibrations of Cu atoms are present. At the highest frequencies ($\sim$900--1200~cm$^{-1}$), the vibrations of all atoms are present. In the soft mode, we observe the dominant contribution from oxygen atoms.

We investigate the stability of the $\beta$ phase at higher temperatures by calculating the phonon dispersion relations using the SCPH procedure. The potential energy was obtained within the TDEP approach up to the quartic terms using the atomic displacements generated by the HECSS method. In calculations, we used $N=100$ configurations of atomic displacements in the supercell with 88 atoms at $T=500$~K. The SCPH equation was solved using the $4\times4\times4$ $q$-point grid. Due to the thermal effects, the soft mode at the A point is stabilized, and its frequency strongly depends on temperature. Its value goes to zero when the temperature is decreased. This clear dependence of phonon frequency on temperature indicates a strong anharmonicity of the soft mode.
Also, the lowest phonon modes at the V, L, and M points show a strong shift to higher frequencies when temperature is increased. This demonstrates the crucial role of anharmonicity in the stabilization of the high-temperature $\beta$ phase of copper pyrophosphate.

\begin{figure}[t!]
  \centering
    	\includegraphics[width=8.6cm]{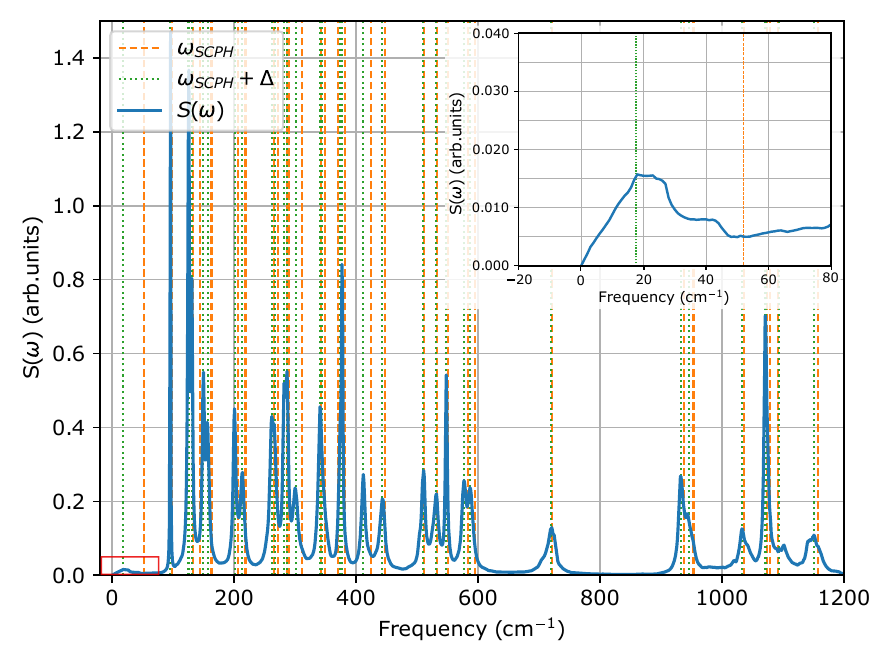}
    \caption{The phonon spectral function obtained from the anharmonic self-energy at the A point with $\bm{q}=(0,0,0.5)$. The positions of phonon peaks obtained within the SCPH method and with the self-energy correction ($\Delta$) are shown by the dashed orange and dotted green lines, respectively.}  
  \label{fig:profile}
\end{figure}

In the final step, we calculate the phonon spectral function $S(\omega)$ using the phonon line width ($\Gamma$)
and frequency shift ($\Delta$) obtained from the phonon self-energy including the bubble diagrams~\cite{Tadano2018}. The calculation of $S(\omega)$ is based on the SCPH results, therefore,
the phonon frequencies are equal to $\omega_\text{SCPH}+\Delta$. In order to study the anharmonic behavior of the soft mode, in Fig.~\ref{fig:profile} we present the phonon profiles obtained for the A point with $\bm{q}=(0,0,0.5)$.
Most of the phonon peaks depicted in Fig.~\ref{fig:profile} show very weak frequency shifts compared to the SCPH values, and their line widths are rather narrow. A radically different behavior is observed for the lowest mode, which exhibits strong damping and softening due to the anharmonic effects.
The shape of the soft-mode profile is very irregular and asymmetric (see the inset in Fig.~\ref{fig:profile}).

\begin{figure}[t!]
  \centering
    	\includegraphics[width=8.6cm]{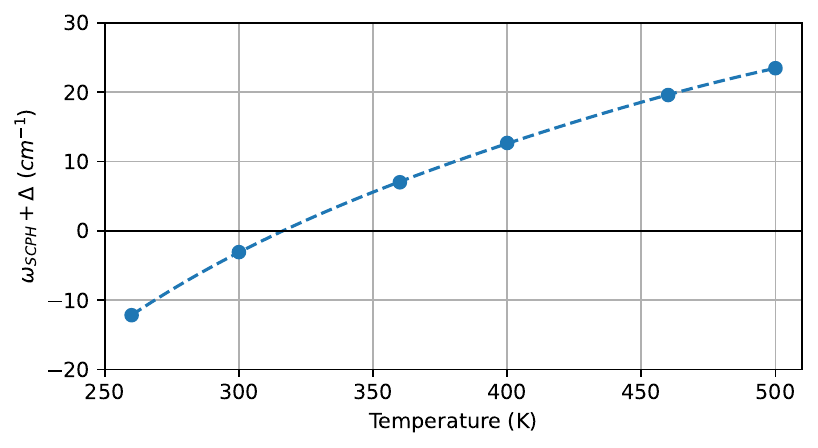}
    \caption{Temperature dependence of the soft-mode frequency.}  
  \label{fig:delta}
\end{figure}

In Fig.~\ref{fig:delta}, we plot the frequency of the soft mode as a function of temperature.
The frequency goes to zero when we approach $T\sim 320$~K, and we can treat it as the temperature of the structural phase transition. It is only the approximate estimation of the critical temperature, but it corresponds quite well to the experimental value. 

In order to better understand the structural transformation, we analyze how
the lattice distortion induced by the soft mode modifies the total energy and position of the O1 atom.
Fig.~\ref{fig:phonons}(c) presents the change of total energy $\Delta E$ calculated per one primitive cell of the $\beta$ phase 
as a function of the displacement $u$ of the O1 atom, obtained using the polarization vector of the soft mode.
The central position of O1 corresponds to $u = 0$. 
The energy decreases, reaching the minimal value for the amplitude $u=0.23$~\AA,
which agrees quite well with the position of O1 found by the diffraction measurements for the $\beta$ phase $u_\text{exp}(\beta)=0.27$~\AA.
It demonstrates that the calculated double-well potential corresponding to the soft mode
determines the shifts of O1 atoms in the high-temperature disordered phase.

Relaxation of the distorted structure leads to lower energy, which is marked in Fig.~\ref{fig:phonons}(c) by the dashed violet line.
However, this value is higher than the ground state energy obtained by the full optimization starting from the experimental
values of atomic positions and lattice parameters within the $\alpha$ phase (dashed blue line).
The displacement of O1 from the central position found for the $\alpha$ phase equals to $u_\text{th}(\alpha)=0.34$~\AA,
and it corresponds well to the experimental value, $u_\text{exp}(\alpha)=0.32$~\AA.
  
Below the phase transition, the soft mode is stabilized and changes into the optical mode at the $\Gamma$ point. 
It is the lowest Raman mode of the $A_g$ symmetry and frequency 87.06~cm$^{-1}$.
This particular mode was previously connected with the strong NTE observed in Cu$_2$P$_2$O$_7$~\cite{Shi2020}.
The transverse motion of O1 atoms triggers the twist and rotation of the CuO$_5$ and PO$_4$ polyhedra, which were recognized as the inherent factors for the NTE.
The diffraction studies revealed that the lattice parameters $a$ and $c$ decrease with increasing temperature, up to the $\alpha$-$\beta$ phase transition.

\section{CONCLUSIONS}

In summary, the electronic and lattice properties of the $\beta$-Cu$_2$P$_2$O$_7$ were investigated by {\it ab initio} calculations. 
The obtained results were compared with the previous studies performed
for the $\alpha$ phase~\cite{Pastukh2021} and the available experimental data.
For both phases, the lattice parameters show a good agreement with the experimental data if local Coulomb interactions and van der Waals corrections are included in the calculations. 
The Mott insulating band gap of $E_g=2.94$~eV is slightly smaller than in the $\alpha$ phase (3.17~eV) and corresponds well to the experimental data~\cite{karaphun2018influence,Majtyka2022}.
The AFM configurations found in both phases, with the magnetic moments equal to 0.86~$\mu_\text{B}$, are very similar and agree with the long-range order observed experimentally below $T_\text{N}=27$~K.
Based on phonon dispersion relations and PDOS, the dynamical properties of the $\beta$-Cu$_2$P$_2$O$_7$ crystal were analyzed.
The soft mode discovered at the A point indicates the dynamical instability of the $\beta$ phase at low temperatures.  It is responsible for the structural transition, in which the symmetry changes from the monoclinic $\alpha$ phase ($C2/c$) to the monoclinic $\beta$ phase ($C2/m$).
The soft mode induces primarily vibrations of O1 atoms, resulting in static displacements along the $b$ direction and doubling of the lattice parameter $c$ in the $\alpha$ phase. 
The obtained double-well potential allowed us to estimate the positions of O1 atoms in a good correspondence to the diffraction data.
The calculations performed within the TDEP+SCPH method demonstrated the strong temperature dependence of the soft mode, which leads to the stabilization of the $\beta$ phase at higher temperatures. The transition temperature was estimated to 320 K, close to the experimental value of approximately 360~K~\cite{Shi2020,Robertson1967,Robertson1968,Pogorzelec2006}. Strong anharmonicity of the soft mode is additionally confirmed by the large phonon linewidth obtained from the phonon self-energy.
As it was previously observed, the NTE is related to the same vibrations of the O1 atoms along the direction perpendicular to the P--O--P bridge. Our analysis indicates that the lowest optical $A_g$ mode, which is responsible for the NTE, corresponds to the soft mode found in the $\beta$ phase. These observations may be important for further investigations of structural phase transitions and NTE in other pyrophosphate materials.

The presented results demonstrate that the anharmonic and thermodynamic properties of complex materials can be successfully studied within the TDEP and SCPH methods in connection with the recently developed HECSS approach~\cite{Jochym2021}.
It opens another route to investigate even more complicated materials, molecular crystals, or heterostructures.

\section*{ACKNOWLEDGMENTS}

This work was supported by the Ministry of Education, Youth and Sports of the Czech Republic through the e-INFRA CZ (ID:90254) project and project No. LUASK22099. P.P. acknowledges the support of the National Science Centre (NCN, Poland) under Project No. 2021/43/B/ST3/02166. S.P. acknowledges financial support provided by the Polish National Agency for Academic Exchange NAWA  under the Programme STER– Internationalisation of doctoral schools, Project no.  PPI/STE/2020/1/00020.

\bibliography{biblio}

\end{document}